\documentclass[prb,letterpaper,aps,floatfix,twocolumn]{revtex4}
\usepackage{graphicx}
\usepackage{amsmath}
\newcommand{\beq}{\begin{equation}}
\newcommand{\eeq}{\end{equation}}
\newcommand{\bk}{{{\bf{k}}}}

\newcommand{\br}{{{\bf{r}}}}

\newcommand{\bA}{{\bf{A}}}

\newcommand{\beqa}{\begin{eqnarray}}
\newcommand{\eeqa}{\end{eqnarray}}

\newcommand{\dg}{{\dag}}
\newcommand{\pdg}{{\vphantom\dag}}
\newcommand{\btau}{{\boldsymbol \tau}}
\newcommand{\bsigma}{{\boldsymbol \sigma}}
\begin{document}
\title{Parallel magnetic field driven quantum phase transition in a thin topological insulator film}
\author{A.A. Zyuzin, M.D. Hook, and A.A. Burkov}
\affiliation{Department of Physics and Astronomy, University of Waterloo, Waterloo, Ontario 
N2L 3G1, Canada}
\date{\today}
\begin{abstract}
It is well-known that helical surface states of a three-dimensional topological insulator (TI) do 
not respond to a static in-plane magnetic field. Formally this occurs because the 
in-plane magnetic field appears as a vector potential in the Dirac Hamiltonian of the 
surface states and can thus be removed by a gauge transformation of the 
surface electron wavefunctions. Here we show that when the top and bottom surfaces 
of a thin film of TI are hybridized and the Fermi level is in the hybridization gap, a nonzero {\em diamagnetic} response appears.
Moreover, a quantum phase transition occurs at a finite critical value of the parallel field from an insulator with a diamagnetic response to a semimetal with a vanishing response to the parallel field.
\end{abstract}
\maketitle
\section{Introduction}
Topological insulators (TI) are long-known and relatively common materials, which have recently been shown to exhibit phenomena, closely related to the 
quantum Hall effect (QHE), but  under significantly less stringent conditions, requiring neither the two-dimensionality, nor the high magnetic fields or the ultralow temperatures, necessary to observe the QHE.~\cite{Kane10}
In particular, TI materials, under the condition of unbroken time-reversal symmetry (TRS), have robust metallic edge states. 
These edge states are not chiral, like in the QHE case, but helical, which means that for each momentum on the Fermi surface of the edge state, the spin has a rigidly defined direction, transverse to the momentum.~\cite{Kane05,Hasan09}
The spin direction thus rotates by $2\pi$ around the Fermi surface, which can be thought of as a surface manifestation of the nontrivial topology of the bulk bandstructure of the material. 
This rigid spin-momentum coupling is naturally viewed as a very interesting property of the TI surface states from the point of view of spintronic 
applications.~\cite{Raghu10,Garate10,Sengupta10,Nagaosa10,Burkov10,DasSarma10}
Generation of current with 100\% spin polarization (giant inverse spin-galvanic effect)~\cite{Garate10,Nagaosa10} and novel 
magnetoresistance effects~\cite{Burkov10} are some of the concrete proposals that have been discussed in theoretical literature. 
One practical obstacle that currently stands in the way of realizing these proposals experimentally, is the large bulk conductivity that most 
known TI materials possess due to unavoidable imperfections in the crystal structure and composition. 
While different methods of dealing with this problem have been proposed, perhaps the most attractive one is to grow TI samples in the form 
of ultrathin films, so that the bulk conductivity contribution, relative to that of the surface states, is naturally reduced and does not mask the 
surface transport effects.~\cite{ZhangG09}

Apart from resolving a practical problem, ultrathin TI films have interesting physical properties of their own, that bulk samples do not have. 
Nontrivial collective effects, such as excitonic superfluidity,~\cite{Seradjeh09} have been theoretically proposed to occur in 
ultrathin TI films. 
Magneto-optical response of thin TI films has been predicted to possess universal properties, reflecting topological order of the 
bulk material.~\cite{Tse10} 
Yet another group of interesting phenomena in TI films occur when the thickness of the film becomes comparable to the ``penetration depth" of the helical surface states into the bulk and the top and bottom surfaces thus start to hybridize. This typically happens at a thickness of 5 to 10 quintuple layers (QL), i.e. of the order of 10 nm.~\cite{ZhangG09} 
In particular, if the Fermi level is situated within the hybridization gap, the film may (depending on the sign of the hybridization matrix element) 
enter the two-dimensional quantum spin Hall insulator phase,~\cite{Liu10} or the quantum anomalous Hall insulator phase if TRS 
is broken.~\cite{Yu10} 

In this paper we show that a thin TI film with hybridization between the helical metal states on opposite surfaces possesses a highly nontrivial 
response to magnetic field, applied in the plane of the film. It is well-known that TI surface states respond strongly to a perpendicular magnetic field: the gaplessness of the 
surface Dirac spectrum is guaranteed by TRS, but once TRS is broken by the perpendicular field, a gap opens up at the Dirac point. 
If the Fermi level is in the gap, the surface becomes insulating and exhibits half-integer-quantized Hall effect.~\cite{Kane10} 
Response to parallel field, however, is completely different. While it is known that an isolated helical surface state does not respond to the parallel field, 
apart from a rigid shift of the Dirac dispersion in momentum space, this changes, as we demonstrate below, when the surface states on the opposite 
surfaces of the film are hybridized. We demonstrate that in this case the film undergoes a quantum phase transition, driven by the parallel field, 
from an insulating to a semimetallic state. The two states can be distinguished by their magnetic response: while the insulator has a diamagnetic response, 
the response in the semimetal is identically zero (this does not include the magnetic response of the bulk TI material, see below). 
We note, in passing, that parallel field response of the helical edge states has also been discussed for two-dimensional TI samples in the form of a thin strip.~\cite{Zhou08} While there are some similarities, the phase transition we discuss below is absent in this case (it is replaced by a 
symmetry-related band crossing, with no nonanalytic behavior of either the band dispersion or the magnetic susceptibility).  

\section{Field-driven insulator to semimetal phase transition}
We consider a thin TI film, subject to an in-plane magnetic field of magnitude $B$. We assume 
for concreteness that the field is directed along the $x$-axis, with the $z$-axis normal to the plane of the film. 
Choosing Landau gauge for the vector potential $\bA = - \hat y B z$ and assuming that 
$z = \pm d/2$ for the top and bottom surfaces of the film correspondingly, we can write down the 
Hamiltonian for this system in the following form:
\beq
\label{eq:1}
H = \sum_\bk \left[ v_F \tau^z \left(\hat z \times \bsigma \right) \cdot \left(\bk - \kappa_B \tau^z \hat y \right) + 
\Delta \tau^x \right] c^\dg_\bk c^\pdg_\bk. 
\eeq
Here $v_F$ is the Fermi velocity, characterizing the surface Dirac dispersion, 
$\Delta$ is the tunneling matrix element between the top and bottom surfaces~\cite{footnote}, 
$\kappa_B = d/2 \ell^2- 1/2 m v_F \ell^2$ is the magnetic wavevector, which contains contributions from the Aharonov-Bohm phase gradient
(first term) and the Zeeman coupling (second term) and $\ell = \sqrt{c/eB}$ is the magnetic length (we will be using $\hbar =1$ units).
For film thicknesses greater than 1 QL,  $d > 1/ m v_F \sim 1~\textrm{nm}$, and we will neglect the Zeeman contribution to $\kappa_B$ 
henceforth. 
We have introduced two sets of Pauli matrices, $\bsigma$ and $\btau$, acting on the spin 
and the top and bottom surface pseudospin degrees of freedom correspondingly.  The spin and pseudospin 
indices are suppressed for brevity. 

It is well-known that in-plane magnetic field has no effect on helical surface electrons, since the corresponding term in the Hamiltonian can be eliminated by 
a gauge transformation of the electron field operators $\Psi(\br) \rightarrow \Psi(\br) e^{i \tau^z \kappa_B y}$, 
which simply shifts the surface Dirac cones by $\pm \kappa_B$ along the $y$-axis in momentum space. However, since the corresponding phase 
factors are complex conjugates of each other for the opposite surfaces of the film, the tunneling term in Eq.(\ref{eq:1}) breaks this ``gauge symmetry"
(this is known as {\em chiral symmetry} in particle physics) and thus leads to a nontrivial dependence of the energy of the system on the 
parallel field. 

\begin{figure}[t]
\includegraphics[width=9cm]{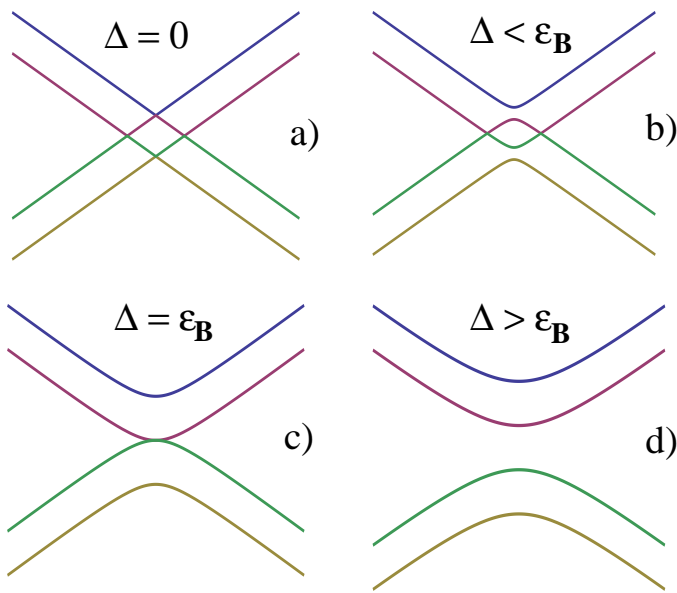}
\caption{(Color online) $k_x = 0$ section of the surface band dispersion, plotted at different values 
of the tunneling matrix element $\Delta$ and fixed nonzero magnetic field.
(a) $\Delta = 0$ bandstructure: two Dirac cones shifted by $2 \kappa_B$ with respect to each other 
along the $y$-axis.
(b) Bandstructure for $\Delta < \epsilon_B$, i.e. in the semimetal phase. Hybridization gaps open only at the two $k_y = 0$ crossing points. 
(c) Critical point $\Delta = \epsilon_B$. The dispersion of the two bands that touch is quadratic
along the $y$-direction, but remains linear in the $x$-direction. 
(d) Bandstructure in the insulating phase, $\Delta > \epsilon_B$. } 
\label{fig:1}
\end{figure}

To see this explicitly, we diagonalize the Hamiltonian Eq.(\ref{eq:1}), obtaining the following surface band dispersion:
\beq
\label{eq:3}
\epsilon_{\bk \pm} = \sqrt{v_F^2 \bk^2 + \Delta^2 + \epsilon_B^2 \pm 2 \epsilon_B\sqrt{v_F^2 k_y^2 + \Delta^2}},   
\eeq
where $\epsilon_B = v_F \kappa_B$ is the ``magnetic energy" and the second pair of bands is given by $-\epsilon_{\bk \pm}$. 
The most interesting feature of this bandstructure is a quantum phase transition, that occurs as a function of $\epsilon_B/\Delta$. 
This transition is illustrated in Fig.~\ref{fig:1}.
Indeed, when $\epsilon_B < \Delta$ and focusing on the positive energy part of the spectrum, the bandstructure consists of a pair of quadratically dispersing 
bands, split by the TRS-breaking gap of magnitude $2 \epsilon_B$ at $\bk = 0$. The bottom of the lower band is above the Fermi energy $\epsilon_F = 0$ 
by $\Delta - \epsilon_B$ and the film is thus an insulator with an energy gap of $2 (\Delta - \epsilon_B)$. 
When $\epsilon_B = \Delta$, the gap closes and the system becomes a semimetal. Right at the transition point, the conduction and the valence bands 
touch at a single point, which has an unusual character: the band dispersion away from this point is linear in the $x$-direction (i.e. along the magnetic field) and quadratic in the $y$-direction. Such ``semi-Dirac" band-touching points also occur (via a different mechanism) in 
graphene~\cite{semidirac} and have interesting properties. 
Here we will, however, focus on the physics of the field-driven semimetal-insulator transition rather than on the properties of the critical point itself. 
When $\epsilon_B > \Delta$, the film is semimetallic and the low-energy part of the dispersion consists of two Dirac points, separated by a wavevector
$2 \kappa_0 = (2/ v_F) \sqrt{\epsilon_B^2 - \Delta^2}$ along the $y$-axis in momentum space. 

To understand the nature of this phase transition better, it is useful to approach it from the opposite limit, namely the limit of zero hybridization $\Delta=0$. 
In this limit, the bandstructure consists of two Dirac cones, shifted by $2 \kappa_B$ along the $y$-axis, as illustrated in Fig.~\ref{fig:1}.
Focusing on the $k_x=0$ section of the band dispersion, shown in Fig.~\ref{fig:1},  there are four band-crossing points: two corresponding to the 
Dirac points at $k_y = \pm \kappa_B$ and two at $k_y = 0$.
Turning on a small intersurface tunneling matrix element $\Delta$, one would expect, naively, hybridization gaps to open up at all four crossing points. 
This, however, is not what happens. Gaps of magnitude $2 \Delta$ in fact open up only at the two crossing points at $k_y = 0$, {\em but not at the Dirac 
points}. The reason for  this is topological, as was noted in a different context by Klinkhamer and Volovik.~\cite{Volovik05} 
Each Dirac point can be characterized by a topological 
invariant: the spin winding number around any surface in momentum space, enclosing the point (the winding numbers have opposite signs for the positive and negative energy bands, emanating from each Dirac point).  The winding numbers are equal to unity by magnitude 
and have opposite signs for the two Dirac points. As long as the Dirac points are separated in momentum space, they are stable with respect to 
the intersurface hybridization, since only states in momentum space with the same spin can be hybridized. Thus the Dirac points can not be eliminated by 
turning on a small tunneling matrix element in the Hamiltonian.
The pair of Dirac points can only be annihilated (like a vortex-antivortex pair) by merging the points together, which is precisely what happens 
when $\epsilon_B = \Delta$. The parallel-field driven semimetal-insulator transition in a thin TI film is thus a topological transition, in the sense that the two sides of the transition are distinguished by the presence or absence of separated neutral pairs of topological defects in momentum space, i.e. the Dirac points. 
An important caveat one needs to keep in mind is that once an in-plane magnetic field is applied to the film, the Dirac points are no longer protected against disorder. Even though the in-plane field does not open a gap, it moves the Dirac cones to different points in momentum space, as discussed above. Any translational symmetry breaking perturbation can then hybridize the Dirac cones, if it has a Fourier
component at the wavevector, connecting them. Thus, strictly speaking,  Dirac points continue to be topologically protected 
only if translational symmetry is imposed. However, we expect that our results will not change  significantly as long as the disorder scattering is weak, namely as long as $\Delta \tau \gg 1$, where $\tau$ is the mean scattering time.  

In the remaining part of the paper we will demonstrate that the two phases, the semimetal and the insulator, can be sharply distinguished 
by their magnetic response to the in-plane field: while the insulator has a nonvanishing diamagnetic response, the response of the semimetal is identically zero. The magnetic susceptibility $\chi$ can thus be used as an ``order parameter", characterizing the above phase transition. 

Magnetic susceptibility can be calculated as the second derivative of the free energy of the system:
\beqa
\label{eq:4}
F&=&-T \sum_\bk \left[ \ln\left( 1 + e^{- \frac{\epsilon_{\bk +} - \epsilon_F}{T}} \right) + 
\ln\left( 1 + e^{- \frac{\epsilon_{\bk -} - \epsilon_F}{T}} \right) \right. \nonumber \\
&+&\left. \ln\left( 1 + e^{\frac{\epsilon_{\bk +} + \epsilon_F}{T}} \right) +  \ln\left( 1 + e^{\frac{\epsilon_{\bk -} + \epsilon_F}{T}} \right) \right],
\eeqa
with respect to the magnetic field $\chi(B) = - (1/ d L^2) \partial^2 F / \partial B^2$, 
where $L$ is the linear dimension of the film in the $x,y$-directions.
A straightforward calculation assuming $T= 0$ and $\epsilon_F = 0$ gives $\chi(B) = (v_F d e / 2 c)^2 {\cal I}(B)$,
where ${\cal I}(B)$ is the following momentum-space integral, which includes the contribution of all filled (i.e. negative-energy since we take 
$\epsilon_F = 0)$ single-particle states:
\beqa
\label{eq:7}
{\cal I}(B)&=&\int \frac{d^2 k}{(2 \pi)^2} v_F^2 k_x^2 \nonumber \\ 
&\times&\left[ \frac{1}{\epsilon_{\bk +}^3} + \frac{1}{\epsilon_{\bk -}^3} - \left. \frac{1}{\epsilon_{\bk +}^3} \right|_{\Delta = 0} - 
\left. \frac{1}{\epsilon_{\bk -}^3} \right|_{\Delta = 0} \right]. 
\eeqa
The last two terms in the square brackets in Eq.(\ref{eq:7}) have been added by hand to remove the ultraviolet divergence in ${\cal I}$, which is a known feature of  Dirac fermions. The subtracted terms correspond to the magnetic susceptibility calculated at $\Delta=0$, which is equal to zero from the discussion above. 

The two-dimensional momentum integral in Eq.(\ref{eq:7}) appears to be rather formidable. Quite unexpectedly, however, it can be evaluated exactly and gives a very simple 
final answer. Indeed, integrating over $k_x$ first, we obtain a significantly simplified expression:
\beq
\label{eq:8}
{\cal I}(B) = \frac{1}{2 \pi^2 v_F} \int_{-\infty}^{\infty} d k_y \ln \left| \frac{k_y^2 - \kappa_B^2}{k_y^2 - \kappa_0^2} \right|. 
\eeq
The integral in Eq.(\ref{eq:8}) behaves very differently depending on the presence or absence of Dirac points in the spectrum, i.e. 
in the insulator and semimetal phases. In the semimetal the wavevector $\kappa_0 = (1/v_F) \sqrt{\epsilon_B^2 - \Delta^2}$, which determines 
the location of Dirac points on the $y$-axis,  is real. The integrand in Eq.(\ref{eq:8}) then has logarithmic singularities that all lie on the real axis.  As a result, the integral is easily shown to evaluate to zero. 
In contrast, as one enters the insulating phase, the Dirac points merge and the insulating gap opens up. The wavevector $\kappa_0$ now becomes imaginary 
and two of the logarithmic singularities in Eq.(\ref{eq:8}) move to the imaginary axis. 
The integral is then nonzero and is given by ${\cal I}(B) = - (1/ \pi v_F^2) \sqrt{\Delta^2 - \epsilon_B^2}$, 
i.e. it is proportional to the separation between the two logarithmic singularities of the integrand in Eq.(\ref{eq:8}) on the imaginary axis.
Thus we obtain the following result for the magnetic susceptibility:
\beq
\label{eq:10}
\chi(B) = - \Theta(\Delta - \epsilon_B) \frac{\alpha^2 m a_0 d}{4 \pi} \sqrt{\Delta^2 - \epsilon_B^2}.
\eeq
Here $\Theta(x)$ is the step function, $\alpha = e^2/c$ is the fine structure constant and $a_0 = 1/m e^2$ is the Bohr radius. 
Susceptibility $\chi(B)$ is {\em diamagnetic} in the insulating phase and is {\em identically zero} in the semimetal. 
The diamagnetic response of the insulator arises due to circulating persistent currents, which flow in the opposite directions on the 
top and bottom surfaces of the film to partially screen the applied in-plane field. 
Note once again, that, as directly follows from the properties of Eq.(\ref{eq:8}), whether the magnetic susceptibility is zero or not, 
depends only on the presence or absence of Dirac points in the spectrum, i.e. on its topological properties. 

Integrating $\chi(B)$, we can also obtain the magnetization density as a function of the field:
\beqa
\label{eq:11}
&&M(B)= - \Theta(\epsilon_B - \Delta - \eta) \frac{\mu_B m \Delta^2}{4 v_F}  - 
\Theta(\Delta - \epsilon_B + \eta) \nonumber \\
&\times&\frac{\mu_B m \Delta^2}{2 \pi v_F} 
\left[\frac{\epsilon_B}{\Delta} \sqrt{1 - \frac{\epsilon_B^2}{\Delta^2}}+\arctan \left(\frac{\epsilon_B}{\sqrt{\Delta^2-\epsilon_B^2}}\right)\right], \nonumber \\
\eeqa
where $\eta = 0+$. 
The semimetal is thus characterized by a constant, field-independent magnetization, which is proportional to the square 
of the hybridization matrix element. 
We would like to point out that the above results are valid as long as the surface state dispersion can be well approximated 
by the linear Dirac dispersion, i.e. as long as the relevant energy scales $\Delta$ and $\epsilon_B$ are significantly 
smaller than the bulk band gap. 

\section{Discussion and conclusions}

Let us now comment on the experimental observability of the proposed phase transition and the nonanalytic behavior
of the magnetic susceptibility $\chi(B)$ which accompanies it. 
First, let us estimate the field strengths that would be necessary to observe the transition. We take the tunneling gap to be                                    $\Delta \sim 1~\textrm{meV}$
and $d \sim 10~\textrm{nm}$. Using $v_F \sim 10^7~\textrm{cm/s}$, we obtain the critical magnetic field to be of the order of 
$1~\textrm{Tesla}$. 
The $1~\textrm{meV}$ hybridization gap also requires correspondingly low, i.e. $\lesssim 10~\textrm{K}$ temperatures.  The above conditions on the 
magnitude of the magnetic field and temperature are relatively easily realizable experimentally. 
Another important issue is the magnitude of the magnetic response under these conditions. Assuming as above that $\Delta \sim 1~\textrm{meV}$, 
we obtain $\chi \sim 10^{-8}$ at $B = 0$. Thus $\chi$ under the above conditions is approximately one to two orders of magnitude smaller than typical electronic magnetic susceptibilities of common metals. Measuring susceptibility of this magnitude is by itself straightforward, 
however one needs to be able to extract it from the total signal, which will contain a significantly larger diamagnetic contribution from the 
bulk TI material. It should be possible to do it by differentiating the measured susceptibility with respect to $B$, since the bulk 
atomic diamagnetic susceptibility can be expected to be constant for fields of magnitude of a few Tesla. 
Probably the most difficult experimental issue is how to position the Fermi level at the Dirac points with high precision, since as-grown films will 
generally turn out to be doped away from charge-neutrality.~\cite{ZhangG09} 
It should be possible to solve this problem by gating the film.  

In conclusion, we have shown that an ultrathin film of a TI material with hybridization between the top and bottom surfaces exhibits 
a nontrivial response to in-plane magnetic field. Namely, the film undergoes a quantum phase transition from an insulator to a semimetal
at a finite critical value of the field. Coming from the large field (semimetal) side of the phase diagram, the transition can be described as 
annihilation of two Dirac cones, characterized by equal in magnitude but opposite in sign spin winding numbers. 
We have also demonstrated that the insulator and the semimetal can be sharply distinguished by their response to the in-plane magnetic field: 
while the insulator has a diamagnetic response, the magnetic susceptibility of the semimetal is identically zero. To put our results in a broader context, it might be useful to compare in-plane field response of an ultrathin TI film and that of a bilayer 
two-dimensional electron gas in the quantum Hall regime at filling factor $\nu=1$.~\cite{MacDonald04}  In the latter case, a (different) in-plane field driven quantum phase transition is also observed and is understood to be a consequence and one of the smoking-gun features 
of collective phenomena, namely excitonic superfluidity.~\cite{Eisenstein94}
In a TI thin film, a field-driven quantum phase transition arises in the absence of any collective effects, but due to topological order, characterizing the bulk TI material. 

\begin{acknowledgments}
We acknowledge useful discussions with Jan Kycia and Ying Ran. Financial support was provided by the NSERC of Canada and a University of Waterloo start-up grant.
\end{acknowledgments}


\begin{thebibliography}{99}
\bibitem{footnote} Note that $\Delta$ will in general be itself a function of momentum $\bk$. The $\Delta$ that appears in Eq.(\ref{eq:1}) should be understood as the leading-order term in the expansion of the full hybridization matrix element with respect to $\bk$. 
This is expected to be a good approximation at energies, low compared to the bulk band gap energy scale.
\end{thebibliography}
\end{document}